\documentclass[twocolumn,showpacs,preprintnumbers,amsmath,amssymb,prc]{revtex4}

\usepackage{graphicx}
\usepackage{dcolumn}
\usepackage{color}
\usepackage{ulem}

\begin{document}

\title{Variational Monte Carlo method for shell-model calculations
  in odd-mass nuclei and restoration of symmetry}

\author{Noritaka Shimizu}
\affiliation{
  Center for Nuclear Study, The University of Tokyo, 
  7-3-1 Hongo, Bunkyo-ku, Tokyo 113-0033, Japan
}
\author{Takahiro Mizusaki}
\affiliation{
  Institute of Natural Sciences, Senshu University, 
  3-8-1 Kanda-Jinbocho, Chiyoda-ku, Tokyo 101-8425, Japan
}

\date{\today}

\begin{abstract}
We investigate two kinds of extensions for the variational Monte Carlo (VMC) 
method with the Pfaffian in the nuclear shell-model calculations. 
One is the extension to odd-mass nuclei, 
for which we find a new Pfaffian expression of the VMC matrix elements. 
We can, thereby, give a unified VMC framework both for even and odd mass nuclei. 
The other is the extension of the variation after angular-momentum projection. 
We successfully implement the full angular-momentum projected trial state 
into the VMC method, which can provide us with the precise yrast energies. 
We also find a unique characteristic, namely that this angular-momentum projection 
in the VMC can be even ``approximately''  performed. 
This characteristic is useful not only for efficient computation but also 
for precise estimation of the yrast energies through the energy-variance extrapolation.
\end{abstract}


\maketitle

\section{Introduction}
\label{sec:introduction}

Variational Monte Carlo is one of the quantum Monte Carlo methods
to solve quantum many-body problems numerically.
While it is a variational method and
the precision of the approximation depends 
on the quality of the trial wave function and the Hamiltonian, 
it is applicable to any Hamiltonian without 
the notorious sign problem.
Therefore, it has been intensively developed 
in various fields, such as condensed matter physics 
\cite{Tahara, ferrari, MVMC, pandhari}
and nuclear physics \cite{qmc-np, vmcsm1}. 
Especially, the advent of the 
stochastic reconfiguration (SR) method \cite{sorella_sr} 
enables us to use a large number of variational parameters efficiently.
Moreover, as a trial state, a particle-number-projected 
Hartree-Fock-Bogoliubov (HFB) 
wave function can be used 
owing to the Pfaffian, 
  which is known to provide us with the 
  compact and computationally effective wave function 
  \cite{pfaffian-pairing}.
This recent progress broadens the applicability to the 
configuration-space method, such as the Hubbard model.

In nuclear physics, the large-scale shell model (LSSM) calculation
is one of the configuration-space methods and 
a powerful model to describe the nuclear spectroscopic information 
precisely. 
However, the number of the many-body configurations 
which appear in the LSSM tends to be huge, 
and the dimension of the Hamiltonian matrix to be diagonalized 
often surpasses the capability of 
the state-of-the-art supercomputers \cite{caurier_rmp}.
In order to avoid this problem 
and to describe the shell-model wave function
in a sophisticated form, 
the pair-correlated wave function, or 
the HFB-type wave function, was 
suggested in the VAMPIR method \cite{vampir}. 
However, the HFB wave function 
is awkward for treating odd-mass system \cite{bally-odd}. 
We have proposed a new formulation of the 
variational Monte Carlo (VMC) method for shell model calculations
for even-mass nuclei \cite{vmcsm1}
and demonstrated its feasibility 
for LSSM calculations. 

In the present paper, we address two kinds of extensions 
of the previously-presented VMC method.
One extension is to handle odd-mass nuclei in the framework of VMC by 
a new Pfaffian expression. We present the common VMC framework both for even and 
odd mass nuclei.
The other extension is the implementation of the variation after angular-momentum projection. 
Since the atomic nucleus is an isolated system,
the restoration of symmetry is crucial for 
the nuclear structure calculations \cite{ringschuck}.
We successfully implement the  trial state with full angular-momentum projection into the
VMC method.
Unlike other applications of angular-momentum projection, 
we find a unique characteristic that full angular momentum projection in the VMC 
can be performed ``approximately''.
This characteristic is useful not only for efficient computation but also 
for precise estimation of the yrast energies through the energy-variance 
extrapolation.
In the condensed matter physics, the projection method was introduced
in, \textit{e.g.}, Ref.~\cite{mizusaki-imada-pirg-proj} 
and it was also introduced into the VMC 
in Ref.~\cite{Tahara}.
The projection method is well-known, but this implementation of the VMC 
is more flexible than the preceding works.
It may be useful to other fields of physics.
 
This paper is organized as follows: 
Section \ref{sec:framework} is devoted to explaining the theoretical framework
of the VMC method and its extension to odd-mass nuclei. 
The numerical results and ``approximate'' projection are discussed 
in Sect.~\ref{sec:numres}. 
The summary is given in Sect.~\ref{sec:summary}.

\section{formulation of the VMC}
\label{sec:framework}

In this section, we briefly describe the formulation of the VMC. 
We introduce a trial wave function in the subsection \ref{sec:trialwf} 
and describe the way how to stochastically evaluate the energy expectation 
value of the wave function in the framework of  the Monte Carlo 
in subsection \ref{sec:mcmc}.
The restoration of rotational symmetry 
by the projection method in the VMC 
is summarized in subsection \ref{sec:proj}.
The variational parameters are 
determined so that the energy is minimized 
utilizing the SR method, 
the details of which are given in appendix \ref{sec:apdx-sr}.

\subsection{Trial wave function}
\label{sec:trialwf}

As a trial wave function for nuclei with $N$ valence particles 
for the present VMC, 
we take $| \psi \rangle$ as  
\begin{equation}
  |\psi\rangle=GP|\phi\rangle ,
  \label{eq:trial} 
\end{equation}
where the $ | \phi \rangle$ is a pair-correlated wave function 
and $P$ is a projection operator,
both of which are discussed later.
The operator $G$ is the Gutzwiller-like factor as
\begin{equation}
  G=e^{\sum_{i\leq j} \alpha_{ij} n_i n_j }
  \label{Gfactor}
\end{equation} 
where $n_i$ is the number operator of the single-particle orbit $i$ 
and  $\alpha$'s are variational parameters.

For even-mass nuclei, the $|\phi \rangle $ is defined as
\begin{equation}
  \label{eq:even-cwf}
  | \phi \rangle = 
  \left(
    \sum_{kk'} f_{kk^{\prime}}c_{k}^{\dagger}c_{k^{\prime}}^{\dagger}
  \right)^{N/2} |-\rangle
\end{equation}
where $f$ is a skew-symmetric matrix, $f_{kk'} = -f_{k'k}$, 
the matrix elements of which are variational parameters.
The $|-\rangle$ is an inert core and the $c_i^{\dagger}$'s are 
proton or neutron creation operator of the single-particle state $i$.
It corresponds to the number projected Hartree-Fock-Bogoliubov wave function 
\cite{egidoring}, 
which is advantageous for the description of 
pairing correlations. 
Note that this wave function contains
the proton-neutron pairing correlations 
in addition to the proton-proton 
and neutron-neutron pairing correlations, 
while the usual HFB method does not include 
proton-neutron pairing correlations. It plays 
a crucial role in understanding the nuclear structure 
of $N=Z$ nuclei \cite{goodman,hasagawa}.

For odd-mass nuclei,  we extend  the trial wave function
$|\phi \rangle $, which is defined as
\begin{equation}
  \label{eq:wf-odd}
  |\phi \rangle = 
  \left( \sum_l h_l c^\dagger_l \right)
  \left( 
    \sum_{kk'} f_{kk'} c^\dagger_{k} c^\dagger_{k'}
  \right)^{(N-1)/2} | - \rangle , 
\end{equation}
where the $h_l$ are additional variational parameters. 
This form is the simplest for odd-mass nuclei.
Hereafter we discuss the VMC formalism for the odd-mass cases. 
The formulation of the even-mass case 
can be seen in Ref.\cite{vmcsm1} 
and is also obtained easily by omitting 
the terms containing the $h_l$ parameters 
in the following formulations, 
that is,
we can give a unified description with this trial wave function 
for even and odd mass nuclei.

The projection operator $P$ serves to 
restore the rotational symmetry, parity symmetry
and $z$-component of isospin such as  
\begin{equation}
  P  =  P^{T_z} P^{\pi} P^{I}_M 
  \label{Proj_j}
\end{equation}
where $P^{T_z}$, $P^{\pi}$ and  $P^I_M$ are projectors of 
the $z$-component of the isospin, the parity $\pi$, 
and the total angular momentum $(I,M)$, respectively.
The angular momentum operator is decomposed into 
the $\langle J_z\rangle =M$ projection and the rest such as
\begin{equation}
  P_{M}^{I}=P_{M} \tilde{P}_{M}^{I}. 
\end{equation}
where 
\begin{equation}
  \tilde{P}_{M}^{I} \equiv
  \frac{2I+1}{4\pi}
  \sum_{K=-I}^I g_K 
  \int d\gamma d\beta \sin\beta d_{MK}^{I}(\beta)e^{-iK\gamma}
  e^{iJ_y\beta}e^{iJ_z \gamma}.
\end{equation}
The $d_{MK}^{I}(\beta)$ is Wigner's $d$-function
and $g_K$ denotes the $2I+1$ variational parameters.

\subsection{Markov Chain Monte Carlo}
\label{sec:mcmc}

We describe how to estimate 
the energy expectation value of the trial wave function. 
First of all, the projection operator of 
the $z$-component of isospin, parity, and $z$-component of
angular momentum is expressed as a linear combination 
of the complete set in the $m$-scheme basis states as
\begin{eqnarray}
  \label{eq:mproj}
  P^{T_z} P^{\pi} P_M  & = \sum_{m\in \{M^\pi\}} |m\rangle \langle m| .
\end{eqnarray}
where the $m$-scheme basis state $|m\rangle$ is defined as 
\begin{equation}
  \label{eq:mscheme}
  |m\rangle = c^\dagger_{m_1}
  c^\dagger_{m_2} \cdot\cdot\cdot
  c^\dagger_{m_N} |-\rangle 
\end{equation}
which is parametrized by a set of 
occupied single-particle states, 
$m=\{m_1,m_2, \cdot\cdot\cdot, m_N\}$.
The $ \sum_{m \in {M^\pi}} $ denotes the summation of any $|m\rangle$ 
in the subspace with $J_z=M$ and $\pi$-parity. 
It is convenient to take $M=I$, especially for the yrast states.

The energy expectation value is
obtained as 
\begin{eqnarray}
  \langle H \rangle
  &=& \frac{1}{\sum_{m \in {M^\pi}}|\langle m|\psi\rangle|^2 }
      \sum_{m \in {M^\pi}} |\langle m|\psi\rangle|^2 
      \frac{ \langle m|H|\psi\rangle}{\langle m|\psi\rangle }  
  \nonumber \\
  &=& \sum_{m \in {M^\pi}} p(m)  E_l(m)
      \label{eq:eex}
\end{eqnarray}
where $p(m)$ is defined as 
$p(m)=|\langle m|\psi\rangle|^2 / \sum_{m}|\langle m|\psi\rangle|^2 $.
$E_l(m)$ is called the local energy and defined as
\begin{eqnarray}
  E_l(m)
  &=
  & \frac{\langle m|H|\psi\rangle}{\langle m|\psi\rangle }  \\
  & = 
  & \frac{1}{\langle m|\psi\rangle }  
    \sum_{m' \in M^\pi} \langle m|H|m'\rangle
    \langle m'|\psi\rangle
    \nonumber 
\end{eqnarray}
where the matrix $H_{mm'}= \langle m|H| m' \rangle $ is
sparse and the summation concerning $m'$ can be computed 
efficiently since the shell-model Hamiltonian $H$ is a
two-body interaction and has
good parity and rotational symmetries.

The weighted summation $\sum_m p(m)$ in Eq.(\ref{eq:eex}) is 
estimated stochastically 
using the Markov Chain Monte Carlo (MCMC) method
in which $|m\rangle$ walks randomly in the $\{M^\pi\}$ subspace 
obeying the probability $p(m)$.
Such random walker of the $m$ scheme basis state was adopted also 
in Refs.~\cite{vmcsm1,mclanczos,csmc}. 
The energy gradient and the overlap matrix
are also estimated 
stochastically by the SR method.

The overlap between the $m$-scheme basis state
and the $|\psi \rangle$ is shown by
\begin{eqnarray}
  \label{eq:psi-olp}
  \langle m | \psi \rangle 
  &=& G(m)  \langle m | P | \phi \rangle  
\end{eqnarray}
with $ G | m \rangle = G(m) | m \rangle $.
Note that $G$ is a diagonal operator for the 
$m$-scheme basis representation 
and is commutable with the projection operator $P$.
This factor usually accelerates the convergence of the
SR iterations.
While this operator can include many-body correlation beyond
the mean-field and pairing correlations, 
its contribution to the energy gain 
is limited in the case of shell-model calculations.
The projected overlap, $\langle m | P | \phi \rangle$, 
is discussed in the following subsection.

\subsection{Angular-momentum projection}
\label{sec:proj}


The projected overlap $\langle m| P |\phi\rangle$ 
is evaluated as
\begin{eqnarray}
  \label{eq:projmele}
  && \langle m| P|\phi\rangle = \langle m| \tilde{P}^I_M|\phi\rangle
     \\
  & = 
  & \frac{2I+1}{4\pi}\int d(\cos\beta) d\gamma 
    d_{MK}^{I}(\beta) e^{-iK\gamma} 
    \langle m| R(\beta,\gamma) |\phi\rangle
    \nonumber \\
  & \simeq 
  & \frac{2I+1}{4\pi} \sum_K g_K \sum_{a}^{N_z} w^{(z)}_{a} e^{-iK\gamma_a}
    \sum_{b}^{N_y} w^{(y)}_{b} d_{MK}^{I}(\beta_b) 
    \nonumber \\
  && \ \ \ \ \  \langle m| R(\beta_b , \gamma_a) | \phi \rangle , 
    \nonumber
\end{eqnarray}
where the integrals over $\cos\beta$ and $\gamma$ are 
numerically approximated by weighted sums.
The points $(\gamma_b, \beta_a)$ and 
its weight factors ($w^{(z)}_{a}$, $w^{(y)}_{b}$) 
for the integrals 
are determined by the trapezoidal rule for $\gamma$ 
and the Gauss-Legendre quadrature 
for $\beta$ \cite{num_recipe} for efficient computation.
The number of the points for integrals, $N_z$ and $N_y$, 
are usually determined to be large enough to 
evaluate the correct expectation value of $J^2$.
The  numbers are taken typically as ($N_z$, $N_y$) = (32, 16).
The rotation of the correlated-pair wave function 
$|\phi\rangle$ is evaluated as 
\begin{eqnarray}
  \label{eq:rot}
  R(\beta,\gamma)|\phi \rangle 
  &=&  e^{iJ_y\beta_b}e^{iJ_z \gamma_a} | \phi\rangle
      \\
  &=& \left( \sum_l h'_l c^\dagger_l \right)
      \left( 
      \sum_{kk'} f'_{kk'} c^\dagger_{k} c^\dagger_{k'}
      \right)^{(N-1)/2} | - \rangle .
      \nonumber 
\end{eqnarray}
with $h' = Rh$,  $f' = RfR^T$. 
The rotation matrix $R$ is defined as 
$R =e^{J_y \beta_b} e^{J_z \gamma_a}$.
Thus the rotated wave function is kept of the same form
thanks to the Baker-Campbell-Hausdorff theorem \cite{ppnp-mcsm}.

In this paper, we find that the overlap between 
this form of the wave function $|\phi \rangle$ 
and the $m$-scheme basis state can be written 
using the single Pfaffian.  
This is shown in Appendix \ref{sec:apdx-pffafian}. 

The variational parameters $h$, $f$, and $g$ are determined so that 
the energy is minimized utilizing the SR method. In this paper,
we show that the angular-momentum projected energy  can be  minimized in 
the VAP framework of the VMC, 
while the unprojected energy is also minimized to determine the wave function
and the projected energy can be evaluated
in the variation-before-projection (VBP) framework \cite{vmcsm1}.
In the VMC approach, ``unprojected'' means 
without full-angular-momentum projector $\tilde{P}^I_M$, but 
with the $J_z$, parity, and $T_z$ projections.


\section{Numerical Results}
\label{sec:numres}

We discuss the VMC results with variation 
after angular-momentum projection (J-VAP)
in the even-mass case in subsection \ref{sec:vap}, 
and the odd-mass case in subsection \ref{sec:odd}.
The J-VAP calculation can give better yrast energies
than those of our previous paper \cite{vmcsm1}, while
it requires a more substantial computational cost. 
In subsection \ref{sec:appproj},  
the ``approximation'' scheme of angular-momentum projection 
is introduced to reduce the computational cost.
We show that this ``approximation'' scheme   
can give a sequence of wave functions, which can be useful for 
the extrapolation using the energy variance.
With the energy variance extrapolation, the exact 
yrast energies can be estimated beyond the limitation of the trial wave function.

\subsection{Variation after projection for even-mass nuclei}
\label{sec:vap}

In this subsection, we demonstrate the VAP calculation 
with the variation after angular-momentum projection 
of $^{48}$Cr in the $pf$ shell.
The GXPF1A interaction is adopted as an effective 
interaction \cite{gxpf1a}.
For the test of VMC calculation, we use a realistic residual interaction, 
not a schematic interaction so as to properly judge the feasibility
of the method.

\begin{figure}[h]
  \includegraphics[width=8cm]{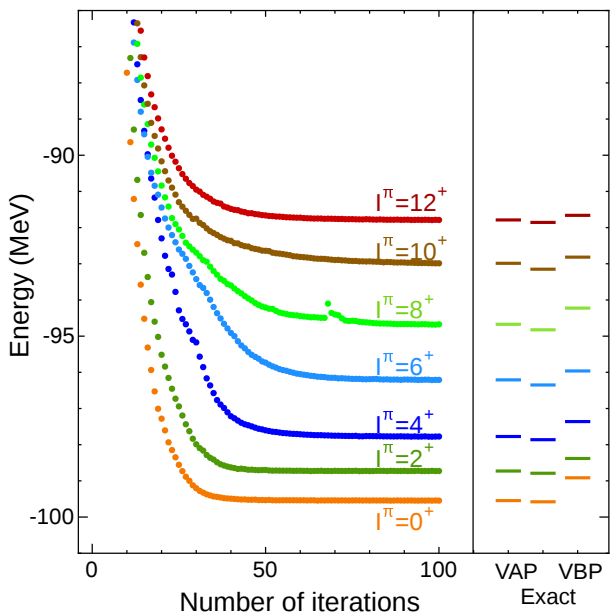}
  \caption{
    Convergence of energies 
    of $I^\pi$ = $0^+, 2^+, 4^+, 6^+, 8^+, 10^+$ 
    and $12^+$ states   of $^{48}$Cr
    obtained by the J-VAP VMC 
    as functions of the number of the SR iterations.
    The right column shows the VAP results, exact shell-model energies,
    and the VBP results.
  }
  \label{fig1}
\end{figure}

Figure \ref{fig1} shows the convergence of the VMC
energy with full angular momentum projection, 
which is called  J-VAP VMC energy later,
as a function of the number of the iterations of 
the SR method. 
The MCMC procedure generates eight random walkers 
with 8000 steps with the Gibbs sampler, the details of 
which are shown in Ref.~\cite{vmcsm1}.
This step needs two-fold integration over Euler's angle 
as in Eq. (\ref{eq:projmele}), which needs heavy numerical computation. 
The present VMC calculations cost a few hours typically on a PC server 
with 56 CPU cores. 
We will show how to reduce the computation later.

The convergence of the J-VAP VMC energies is almost achieved 
with up to 50 $\sim$ 60 steps. 
Since the Monte Carlo error of the energy is typically 
2 keV and small enough, the error bars are omitted for simplicity
in the figure.
The J-VAP VMC energy converges well 
and close enough to the exact shell-model energies 
within 160 keV from $0^+$ to $12^{+}$ states.
For comparison, we show the VBP energy as the rightmost levels
in the figure.
The VMC with VAP improves the energy over VBP as expected.
Especially the VBP result underestimates the $2^+$ excitation 
energy, while  the VAP result sufficiently reproduces
the exact values 
including the backbending phenomenon \cite{ringschuck}; 
{\it e.g.}
$\textrm{Ex}(12^+)-\textrm{Ex}(10^+)$ is smaller 
than $\textrm{Ex}(10^+)-\textrm{Ex}(8^+)$. 
Note that the isoscalar pairing plays an important role 
in the backbending of $^{48}$Cr \cite{pnpair-pf}
and it is shown that the VMC calculations 
are suitable for including the isoscalar-pairing correlations.
The small energy  differences between the exact energies and J-VAP VMC ones
 will be discussed in subsection \ref{sec:energy_variance}.

\subsection{Variation after projection for odd-mass nuclei}
\label{sec:odd}

In this subsection, we consider the odd-mass nuclei for 
a test of the new trial wave function. 
We calculate the yrast energies of 
$^{49}$Cr within the $pf$-shell model space and the GXPF1A interaction \cite{gxpf1a}.
In this VMC calculations, we apply the full angular momentum projection 
to the trial state.
In the MCMC process, we adopt the Gibbs sampler 
with 640 random walkers, each of which 
contains 500 sample steps after 100 burn-in steps.
In order to suppress the biases induced by the initial state 
of the Markov Chain, 
we take the last sample of the previous SR iteration as 
an initial sample of the MCMC process.

\begin{figure}[h]
  \includegraphics[width=8cm]{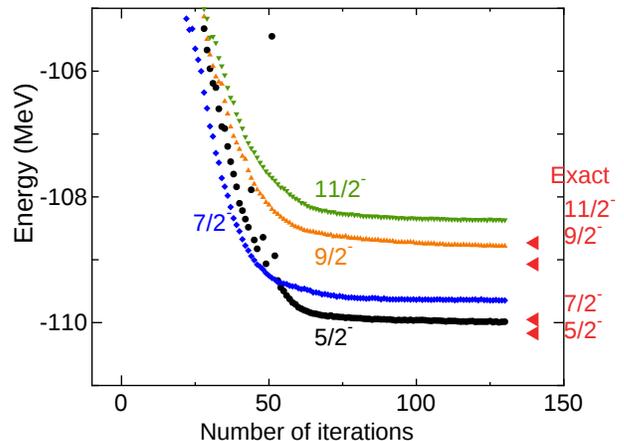}
  \caption{
    Convergence of the J-VAP VMC energies of $^{49}$Cr with 
    GXPF1A interaction.
    The black circles, blue diamonds, orange triangles, 
    and green reverse triangles denote the energy expectation 
    values of $5/2^-$, $7/2^-$, $9/2^-$, $11/2^-$ states, 
    respectively, as functions of the number of iterations.
    The exact shell-model energies are shown as 
    the rightmost red triangles. 
  }
  \label{fig:odd}
\end{figure}

Figure \ref{fig:odd} shows the convergence of the J-VAP VMC energy of 
$^{49}$Cr as an example of odd-mass nuclei.
The energies of the yrast states $5/2^-$, $7/2^-$, $9/2^-$, and $11/2^-$ 
are shown in the figure.
The difference between 
the converged energy and the exact one  is similar to the one of the 
even case, which means that our trial wave function Eq.(\ref{eq:wf-odd})
is considerably more proper. 
However, the number of iterations of the odd case is larger 
than the one of the even case.

\subsection{Approximate angular-momentum projection }
\label{sec:appproj}

Since the correlated-pair wave function $|\psi\rangle$ 
does not have good rotational and parity symmetries, the solution 
spontaneously breaks these symmetries and  
it is crucial to restoring them by the projection method.
In general, the J-VAP has a large effect to minimize the 
energy in the context of the configuration-interaction approach. 
Various variational calculation after the angular-momentum projection 
have been, therefore,  proposed such as 
the Monte Carlo shell model \cite{mcsm}, 
the VAMPIR approach \cite{vampir}, 
and the hybrid multideterminant method \cite{hmd-puddu}. 

In these J-VAP calculations, since the energy and the energy gradient are computed 
under the mathematical conditions 
$[H,P^I_{MK}]=0$ and  $P^I_{ML} P^I_{L'K} = \delta_{LL'} P^I_{MK}$, 
the high-precision 
numerical evaluation 
of the projection is essential.
The insufficient number of points for the integral of the Euler angles 
causes numerical instability and the angular momentum projection fails 
in solving the Hill-Wheeler equation.
The angular-momentum projection is, therefore, a central bottleneck of the computation 
of various variational approaches to the nuclear quantum many-body solver
\cite{hagino-goa}.

On the other hand, in the VMC formalism, since the conditions
$[H,P^I_{MK}]=0$ and  $P^I_{ML} P^I_{L'K} = \delta_{LL'} P^I_{MK}$
are not adopted, 
high precision calculations for the angular-momentum projection 
$\tilde{P}^I_M$ is not necessarily needed, 
which means that the number of mesh points for numerical integration 
could be reduced. 
In fact, 
even when we use a small number of points for the integrals 
and the operator $\tilde{P}^I_M$ is mathematically no longer valid as a projection operator,     
the $\tilde{P}^I_M|\phi\rangle$ works as a trial wave function with ``approximated''
angular momentum,
because this  wave function is simply a superposition of the 
rotated wave functions of $|\phi\rangle$ with appropriate weight coefficients as 
\begin{equation}
  \label{eq:sumproj}
  \tilde{P}^I_M|\phi \rangle 
  \simeq \sum_{a=1}^{N_z} \sum_{b=1}^{N_y}  w_a^{(z)}w_b^{(y)} 
  R(\beta_b,\gamma_a) | \phi \rangle.
\end{equation}
Therefore,  as an approximation to the projection operator,  
we introduce the $\tilde{P}^{I}_{M}$ 
with a set of the small numbers of $N_z$ and $N_y$
and call it $\tilde{P}'^{I}_{M}$ hereafter.
Note that this $\tilde{P}'^I_{M}$  is still commutable 
with the operator $G$ for any $(N_z,N_y)$.

\begin{figure}[h]
  \includegraphics[width=8cm]{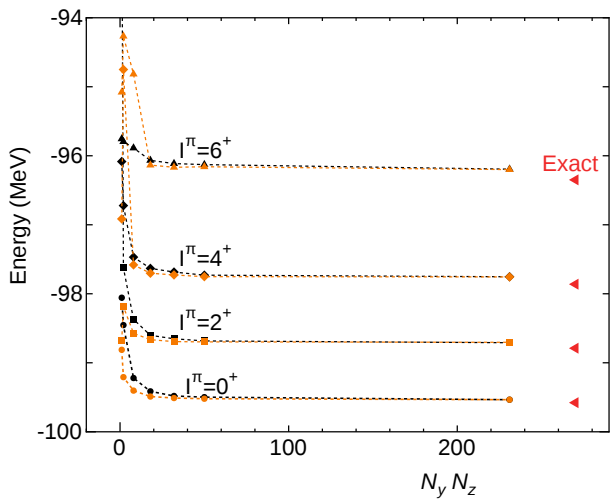}
  \caption{
    VMC results with the variation after approximate angular-momentum 
    projection against the total number of the mesh points 
    for the integral, $N_zN_y$. 
    The black circles, squares, and triangles denote the 
    converged results of the $I^\pi=0^+, 2^+, 4^+,$ and $6^+$ states of $^{48}$Cr, 
    respectively, with the GXPF1A interaction \cite{gxpf1a}.
    The orange symbols denote the full-projected energy of 
    the resultant wave function.
    These symbols are connected with the dotted lines to guide the eyes.
    See text for further details.
  }
  \label{fig:app}
\end{figure}

Figure \ref{fig:app} shows 
the converged VMC energies of the $0^+$, $2^+$ and  $4^+$ energies in $^{48}$Cr
with the GXPF1A interaction \cite{gxpf1a}
as functions of the number of points for the integral of 
the projection operator $\tilde{P}'^I_{M}$.
The VMC calculation was performed with variation 
after the  $\tilde{P}'^I_{M}$ projection.
The number of the points is taken as 
$ (N_z, N_y) = (1,1), (2,1), (4,2), (6,3), (8,4), (10,5)$, and $(21,11)$. 
The converged energies of the variation 
after the approximated projection are shown as the black symbols in 
Fig.~\ref{fig:app}.
The case of $ (N_z, N_y) = (1,1) $ 
corresponds to the variation without the angular-momentum projection. 
In the figure, the rightmost red triangles denote the exact shell-model energies.
The VMC results well reproduce the exact one
even with the small number of $N_zN_y$.
In order to improve the precision of the angular-momentum projection 
so that the expectation value of $J^2$ equals $I(I+1)$ 
exactly to 6 decimal digits, 
the necessary number of points is higher than the minimal one given by 
$(N_z,N_y) = (28,14), (28,14), (31,16),$ and $(35,18)$ 
for $I^\pi = 0^+, 2^+, 4^+,$ and $6^+$ states, respectively.

Astonishingly, the approximated projection works well even 
for $(N_z,N_y)=(6,3)$. 
The total number of the points, $N_zN_y$, is almost proportional 
to the amount of computations of the projected matrix elements, 
which is the most time-consuming part of the VMC calculations.
Therefore, the computation time is dramatically reduced 
in comparison with the full projection.
The required number of the points is rather constant 
as a function of the angular momentum $I$, 
while in the case of the full angular-momentum projection
the necessary number of points increases as $I$ does.
However, we should mention 
that the numerical calculation is stable 
as far as no higher spin state exists 
in the lower-energy region than the target state.

Moreover, we apply the $\tilde{P}^I$ projection, in which the $N_zN_y$ is large 
enough to obtain the correct expectation value of $J^2$ 
to the resultant wave functions.
The orange symbols in Fig.~\ref{fig:app} denote the 
``full'' angular-momentum projected energies.
It is considered to be the variation 
after the approximated projection before the full projection.
These energies are quite close to those of the variation after full projection. 
In practice the energies obtained by the $(N_z,N_y)=(6,3)$ variation 
agree with those of the J-VAP VMC 
within 70-keV difference.

\subsection{Energy-variance extrapolation }
\label{sec:energy_variance}

As the VMC is a variational method, it must not necessarily give us 
exact energies.
The obtained energy is an upper limit.
To know the exact energy, one useful method is 
energy-variance extrapolation 
\cite{sorella_sr, deformed-extrap, imadakashima, mcsm-extrap, mizusaki-ph-extrap},
which uses a series of the well-approximated wave functions 
$|\psi_1 \rangle, |\psi_2 \rangle, \cdots $ 
with monotonically decreasing energies $\langle \psi_1| H |\psi_1 \rangle 
> \langle \psi_2 |H|\psi_2 \rangle > \cdots $.
By  evaluating the energy variance as
$\langle \Delta H^2 \rangle = \langle H^2 \rangle  - \langle H \rangle^2 $
for each wave function, we can show a linear or quadratic relation between the energy variances
and the energies and show that the energy approaches the exact energy along the sequence.
By fitting a second-order polynomial for data points of energy variance and energy, 
the exact energy can be expected by extrapolating the energy to the limit of $\langle \Delta H^2 \rangle=0$.

\begin{figure}[h]
   \includegraphics[width=8cm]{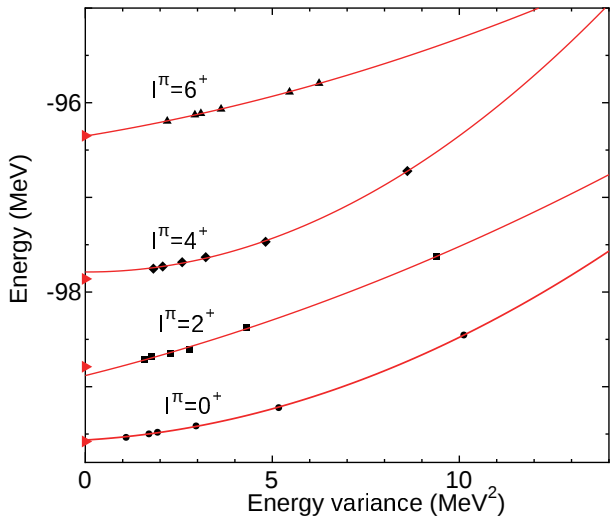}
  \caption{
    Energy variance extrapolation 
    by the variation after the approximate angular-momentum projection.
    The $0^+, 2^+, 4^+,$ and $6^+$  shell-model energies of $^{48}$Cr
    are obtained with the GXPF1A interaction \cite{gxpf1a}.
    The energy expectation values against the energy variance
    are ploted as the black symbols with the approximate 
    projection. The numbers of the points for the projection 
    are taken as $(N_z,N_y) = (2,1), (4,2), (6,3), (8,4), (10,5),$ and 
    $(21,11)$.
    The red lines are chi-square fitted to the symbols. 
    The red squares on the $y$-axis are the exact shell-model energies.
  }
  \label{fig:h2extrap}
\end{figure}

In the preceding application of the energy-variance extrapolation 
to the nuclear shell model,
we used the truncation scheme concerning particle-hole excitations 
to prepare the sequence of well-approximated wave functions \cite{mizusaki-ph-extrap}.
In the present J-VAP VMC scheme, 
the approximate projection method also provides us 
with a sequence of approximated wave functions by changing the number 
of points for the integrals.
This new method can be applied independently of the underlying shell structure.
Figure \ref{fig:h2extrap} shows the 
energy of the VMC with the approximated projection  
as functions of the expectation value of the energy variance, 
$\langle \Delta H^2 \rangle = \langle H^2 \rangle  - \langle H \rangle^2 $.
As the number of mesh points increases, 
the energy expectation values decrease as a function of energy-variances and  
the exact energy is estimated as the intersection of the $y$-axis 
beyond the limitation of the VMC.
These extrapolated energies are close to the exact energies
shown as the red symbols on the $y$ axis.


\section{Summary}
\label{sec:summary}

We presented the VMC method with the Pfaffian to solve the nuclear shell model
in Ref.~\cite{vmcsm1}, where we handle only even-mass nuclei 
and variation before angular-momentum
projection.
In the present paper, we extended the previous VMC method 
for odd-mass nuclei, by deriving 
a new Pfaffian expression for the VMC matrix element. 
We demonstrated that the VMC is successfully applied to 
odd-mass nuclei.
We also extended the VMC to variation after angular-momentum projection,
which enhances the quality of the VMC energy.

In addition to these extensions, we also found that the 
``approximated'' angular-momentum projection can work in the VMC framework.
So far, no feasible approximation scheme for full angular momentum projection 
has been presented, and its numerical calculations have been believed 
to be performed in quite a strict manner. 
However, we proposed a novel approximation scheme of angular momentum projection, 
which reduces the computation drastically and brings about 
an efficient way to calculate angular momentum projection.

Furthermore, we found that this ``approximated'' angular-momentum projection 
also gives a series of well-approximated wave functions, which is useful
to the energy variance extrapolation.
By this development, we could estimate the exact energies of the shell model
beyond the limitation of the VMC.

The form of the trial wave function can be straightforwardly 
extended to that of a one-broken-pair state, which is used 
in Tamm-Dancoff approximation 
and shown in Appendix \ref{sec:apdx-tda}. 
Its numerical application remains as a future subject.

\begin{acknowledgments}
  The authors sincerely acknowledge 
  Professor P. Schuck for carefully reading the manuscript.
  This work was partly supported by KAKENHI grants (25870168, 17K05433)
  from JSPS and priority issue (Elucidation of the 
  fundamental laws and evolution of the universe, 
  hp170230 and hp180179) 
  to be tackled by using Post K Computer from MEXT and JICFuS.
  This work was also partly supported by the research grant of 
  the Senshu Research Abroad Program (2018)
  for one of the authors, (T.M.).
\end{acknowledgments}

\appendix
\section{Pfaffian and its relevant formulas }
\label{sec:apdx-pffafian}

The Pfaffian plays the main role to evaluate 
the matrix elements which appear 
in the present VMC formalism.
Some useful formulas relevant to the Pfaffian are
given in this appendix.
The Pfaffian of a $2n \times 2n$ 
skew-symmetric matrix $A$ is defined as
\begin{eqnarray}
  \label{eq:pfaffian-def}
  \textrm{Pf} (A)  
  & \equiv 
  & \frac{1}{2^{n}n!}\sum_{\sigma\in S_{2n}}
    {\rm sgn}(\sigma)\prod_{i=1}^{n}A_{\sigma(2i-1)\sigma(2i)}  \\
  & = 
  & \frac{1}{n!}\sum_{\sigma\in S_{2n}| \sigma(2i-1)<\sigma(2i) }
    {\rm sgn}(\sigma)\prod_{i=1}^{n}A_{\sigma(2i-1)\sigma(2i)} 
    \nonumber
\end{eqnarray}
where $\sigma$ is a permutation of $\{1,2,3,\cdots , 2n\}$, 
${\rm sgn}(\sigma)$ is its sign, and
$S_{2n}$ is a group of the permutations.

For preparation, the recursive relation of Pfaffian is given as 
\begin{eqnarray}
  \label{eq:recursive_pfaffian}
  \textrm{Pf} (A) 
  &=& \sum_{j=1}^{2n}  (-1)^{i+j+1+\theta(i-j)} A_{ij} 
      \textrm{Pf} (A_{\overline{ij}}) 
\end{eqnarray}
where $A_{\overline{ij}}$ denotes the matrix $A$ 
with the $i$-th and $j$-th columns and rows removed.
$\theta(i-j) $  is the Heaviside step function.
Its special case with $i=1$ is written as 
\begin{eqnarray}
  \label{eq:recur-i1}
  \textrm{Pf} (A) 
  &=& \sum_{j=1}^{2n}  (-1)^{j} A_{1j} 
      \textrm{Pf} (A_{\overline{1j}}) .
\end{eqnarray}

The differentiation of the Pfaffian is given by 
\begin{eqnarray}
  \label{eq:derivpfaff}
  \frac{\partial}{\partial A_{ij}} \textrm{Pf}(A)
  &=& - \textrm{Pf}(A) (A^{-1})_{ij} .
\end{eqnarray}


\section{Overlap with the trial wave function and 
$m$-scheme basis state}
\label{sec:apdx-pair}

In the present VMC formalism, 
the overlap between the trial wave function and the 
$m$-scheme basis state must be computed efficiently.
The trial wave function is a product of the 
Gutzwiller-like operator $G$ and the pair-correlated wave function.
Since the operator $G$ is diagonal in the $m$-scheme basis, 
the overlap is factorized into the matrix element of $G$ 
and the pair-correlated part such as
\begin{eqnarray}
  \label{eq:olpg}
  \langle m | \psi \rangle 
  & = 
  & G(m) \langle m | \phi \rangle 
\end{eqnarray}
with 
\begin{equation}
  G(m) =\exp\left( \sum_{i\leq j} \alpha_{ij} n_i n_j \right)
  \label{Gfactor}
\end{equation} 
where $n_i$ is the number operator of the single-particle orbit $i$ 
and  $\alpha$'s are variational parameters. 
The differential with respect to the variational parameter $\alpha_{ij}$ is 
obtained simply as
\begin{eqnarray}
  \label{eq:gdiff}
  \frac{1}{\langle m | \psi \rangle }
  \frac{\partial}{\partial \alpha_{ij} } 
  \langle m | \psi \rangle 
  & = 
  & n_i n_j .
\end{eqnarray}

The overlap between the pair-correlated wave functions
(\textit{e.g.} Eqs.(\ref{eq:even-cwf}) and (\ref{eq:wf-odd})) 
and $m$-scheme basis state 
in Eq.(\ref{eq:mscheme}) are obtained by using 
the Pfaffian efficiently. Hereafter 
we describe the overlap and its derivative 
concerning the pair-correlated wave functions.

\subsection{Even-mass nuclei}
\label{sec:even-olp}

It is useful to obtain the overlap between 
the $m$-scheme basis state for the $2n$-valence-particles nuclei 
in Eq.(\ref{eq:even-cwf}) 
and the pair-correlated state $|\phi\rangle$.
Using Eq.(\ref{eq:pfaffian-def}), it is obtained as
\begin{equation}
  \label{eq:mphi}
  \langle m| \phi \rangle 
  = \langle m| \left(\sum f_{ij}c^\dagger_i c^\dagger_j\right)^{n}|-\rangle
  = n!\  \textrm{Pf} (F) 
\end{equation}
where $F_{rs} =f_{m_rm_s}-f_{m_s m_r}$. 

Utilizing Eq.(\ref{eq:derivpfaff}), 
its differential is obtained as 
\begin{eqnarray}
  \label{eq:deriv-even-pfaff}
  \frac{1}{\langle m| \psi\rangle} 
  \frac{\partial}{\partial F_{rs} } \langle m| \psi\rangle
  & = & - (F^{-1})_{rs} .
\end{eqnarray}

\subsection{Odd-mass nuclei}
\label{apdx:odd-olp}
The correlated wave function for the odd-mass case 
is defined in Eq.(\ref{eq:wf-odd}).
The number of particles is $N = 2n-1$.
As a novelty, we show the overlap between this odd wave function 
and the $m$-scheme basis state.
Using  Eq.(\ref{eq:recur-i1}), 
the overlap is obtained as
\begin{eqnarray}
  \label{eq:olp-odd}
  \langle m | \phi \rangle 
  & = 
  & \langle m| 
  \left( \sum_l h_l c^\dagger_l \right)
  \left( 
    \sum_{kk'} f_{kk'} c^\dagger_{k} c^\dagger_{k'}
  \right)^{n-1} | - \rangle 
    \nonumber \\
  & = 
  & n!\  \textrm{Pf}(F)
\end{eqnarray}
where $F$ is a $n \times n$ skew-symmetric matrix 
and consists of the first row being $h_{m_p}$
and the other being 
$\tilde{f}_{ij} = f_{m_i,m_j}-f_{m_j,m_i}$
such as 
\newcommand{\tf}[2]{\tilde{f}_{{#1},{#2}}}
\begin{equation}
  \label{eq:tildef}
  F = 
  \left( 
    \begin{array}{cccccc}
      0      & h_{m_1}& h_{m_2} & h_{m_3} & \cdot \cdot \cdot  & h_{m_N} \\
      -h_{m_1}& 0         & \tf{1}{2} & \tf{1}{3} & \cdot \cdot \cdot  & \tf{1}{N} \\
      -h_{m_2}& \tf{2}{1} & 0         & \tf{2}{3} & \cdot \cdot \cdot  & \tf{2}{N} \\
      \cdot \cdot \cdot &&&& \cdot \cdot \cdot \\
      \cdot \cdot \cdot &&&& \cdot \cdot \cdot \\
      -h_{m_N}& \tf{N}{1} & \tf{N}{2} & \tf{N}{3} & \cdot \cdot \cdot  & 0 \\
    \end{array}
  \right) .
\end{equation}

Its differentiation is also obtained in a similar manner
as in the even-mass case such as
\begin{eqnarray}
  \label{eq:deriv-hk}
  \frac{1}{\langle m|\psi\rangle }
  \frac{\partial \langle m|\psi\rangle }{\partial h_{m_k}} 
  &=& - (F^{-1})_{1,k+1} 
      \nonumber \\
  \frac{1}{\langle m|\psi\rangle }
  \frac{\partial \langle m|\psi\rangle }{\partial \tilde{f}_{m_k, m_l}} 
  &=& - (F^{-1})_{k+1,l+1}  .
\end{eqnarray}

\subsection{ Tamm-Dancoff wave function }
\label{sec:apdx-tda}

The wave function used in the Tamm-Dancoff approximation,
which is called a one-broken-pair state, 
is a good approximation to the excited state of the 
pair-condensed wave function in even-mass nuclei having $2n$ 
valence particles.
It can also be used in the VMC formalism, and 
is defined as 
\begin{eqnarray}
  \label{eq:tda}
  | \phi \rangle 
  & = 
  & \left( \sum_{ll'} h_{ll'} c^\dagger_l c^\dagger_{l'} \right)
  \left( 
    \sum_{kk'} f_{kk'} c^\dagger_{k} c^\dagger_{k'}
  \right)^{n-1} | - \rangle .
\end{eqnarray}
Its overlap is obtained using Eq.(\ref{eq:recur-i1}) as
\begin{eqnarray}
  \label{eq:td}
  \langle m |\phi\rangle  
  &=& \sum_{p,q=1}^{2n} (-1)^{p+q-1}h_{m_p m_q} 
      (n-1)!\  \textrm{Pf}(F^{\overline{m_pm_q}})
      \nonumber \\
  &=& \sum_{p=1}^{2n} (-1)^{p} (n-1)!\ 
      \textrm{Pf}(F^{\overline{p}}).
\end{eqnarray}
$F^{\overline{p}}$ is defined as 
$F^{\overline{p}}_{1,1}= 0$, $F^{\overline{p}}_{1,r+1}= h_{m_p,m_r}$, 
and 
$F^{\overline{p}}_{r+1,s+1}= \tilde{f}_{rs} = f_{m_r,m_s}-f_{m_s,m_r}$ with $r,s\neq p$
such as
\begin{equation}
  \label{eq:tildef}
  F^{\overline{p}} = 
  \left( 
    \begin{array}{cccccc}
      0      & h_{m_p,m_1}& h_{m_p,m_2} & h_{m_p,m_3} & \cdot \cdot (p) \cdot  & h_{m_p,m_{2n}} \\
      -h_{m_p,m_1}& 0         & \tf{1}{2} & \tf{1}{3} & \cdot \cdot (p) \cdot  & \tf{1}{2n} \\
      -h_{m_p,m_2}& \tf{2}{1} & 0         & \tf{2}{3} & \cdot \cdot (p) \cdot  & \tf{2}{2n} \\
      \cdot \cdot (p) \cdot &&&& \cdot \cdot (p) \cdot \\
      -h_{m_p,m_N}& \tf{2n}{1} & \tf{2n}{2} & \tf{2n}{3} & \cdot \cdot (p) \cdot  & 0 \\
    \end{array}
  \right) ,
\end{equation}
where $(p)$ denotes that the index $p$ is skipped.
The extension to 
more-broken-pairs states is also expected.

\section{Stochastic Reconfiguration}
\label{sec:apdx-sr}

In the present VMC framework, 
many variables are optimized simultaneously to minimize the 
energy expectation values stochastically.
Although the stochastic estimation of the gradient vector  
enables us to use the steepest gradient method, 
it is unstable due to the stochastic error. 
In order to stabilize the numerical calculation 
and to accelerate it, 
S. Sorrella introduced the stochastic reconfiguration 
(SR) method \cite{sorella_sr}. 
In this appendix, we describe the details of 
the SR method with variation after the angular-momentum 
projection.

The angular-momentum projection 
obliges us to introduce complex numbers as variational parameters, 
while only real numbers are often used as variational parameters 
in the preceding works in condensed matter physics 
(e.g. \cite{Tahara}).
Here, we describe the extension of the SR method of the projected 
wave function including
complex numbers as variational parameters.

We define a derivative operator ${\cal O}_i$, 
which is diagonal in the $m$-scheme basis states, 
and its conjugate operator ${\cal O}^\dagger_i$ such as
\begin{eqnarray}
  {\cal O}_i &=&  \sum_m |m\rangle 
  \left[ \frac{1}{\langle m | \psi_{\bf \alpha} \rangle} 
    \frac{\partial}{\partial \alpha_i} \langle m| \psi_{\bf \alpha} \rangle \right]
  \langle m | 
  \nonumber \\
  &=& \sum_m |m \rangle O_i(m,\alpha) \langle m| 
  \nonumber \\ 
  {\cal O}_i^\dagger &=&  \sum_m |m\rangle 
  \left[ \frac{1}{\langle  \psi_{\bf \alpha} |m \rangle} 
    \frac{\partial}{\partial \alpha^*_i} \langle \psi_{\bf \alpha} | m \rangle \right]
  \langle m | 
  \nonumber \\
  &=& \sum_m |m \rangle O^*_i(m,\alpha) \langle m|, 
\end{eqnarray}
with 
\begin{eqnarray}
  O_i(m,\alpha)&=&  \frac{1}{\langle m | \psi_{\bf \alpha} \rangle} 
  \frac{\partial}{\partial \alpha_i} \langle m| \psi_{\bf \alpha} \rangle 
  \nonumber \\ 
  O^*_i(m,\alpha) &=&  \frac{1}{\langle  \psi_{\bf \alpha} |m \rangle} 
    \frac{\partial}{\partial \alpha^*_i} \langle \psi_{\bf \alpha} | m \rangle , 
\end{eqnarray}
and $\alpha$ denotes a set of variational parameters
 which are complex numbers.
In the present work for the odd-mass case, the variational parameters 
are $\alpha = \{ g_K, \alpha_{ij}, h_l, f_{kk'} \}$.
These operators satisfy the following derivative equations, 
\begin{eqnarray}
  \langle m | {\cal O}_i | \psi_{\bf \alpha} \rangle   
  &=& \frac{\partial}{\partial \alpha_i} \langle m | \psi_{\bf \alpha}\rangle 
  \nonumber \\ 
  \langle \psi_{\bf \alpha} | \hat{O}_i^\dagger | m \rangle   
  &=& \frac{\partial}{\partial \alpha^*_i} \langle  \psi_{\bf \alpha}|m\rangle 
   =   \langle m | {\cal O}_i | \psi_{\bf \alpha} \rangle^* .
\end{eqnarray}


The normalized trial wave function is written as
\begin{equation}
  | \overline{\psi}_{\bf \alpha} \rangle 
  = \frac{1}{\sqrt{\langle \psi_{\bf \alpha}|\psi_{\bf \alpha}\rangle}}
  | \psi_{\bf \alpha} \rangle  .
\end{equation}

The derivative of the normalized trial wave function 
with respect to ${\bf \alpha}$ can be written as 
\begin{eqnarray}
  \frac{\partial}{\partial \alpha_i}  
  | \overline{\psi}_{\bf \alpha} \rangle
  &=& \left( {\cal O}_i - \frac12 \langle {\cal O}_i \rangle \right)
  | \overline{\psi}_{\bf \alpha} \rangle 
  \nonumber \\
  \frac{\partial}{\partial \alpha^*_i}|\overline{\psi}_{\rm \alpha} \rangle
  &=& - \frac12 \langle {\cal O}^\dagger_i \rangle | \overline{\psi}_{\rm \alpha}\rangle
\end{eqnarray}
where we use the shorthand notation $\langle {\cal O} \rangle 
= \langle \overline{\psi} | {\cal O} | \overline{\psi} \rangle$.

The energy gradient $g_i$ is  obtained utilizing 
these derivative operators as
\begin{eqnarray}
  g_i &\equiv&  
  \frac{\partial}{\partial \alpha^*_i} \langle \overline{\psi} |H
  | \overline{\psi} \rangle
  \nonumber \\
  &=& \langle {\cal O}^\dagger_i H \rangle 
  - \langle {\cal O}^\dagger_i\rangle \langle H \rangle , 
\end{eqnarray}

We evaluate $\langle {\cal O}^\dagger_i\rangle$,  $\langle {\cal O}_i\rangle$, 
$\langle {\cal O}^\dagger_i {\cal O}_j\rangle$ and 
$\langle {\cal O}^\dagger_iH \rangle$ stochastically by 
\begin{eqnarray}
  \langle {\cal O}^\dagger_i \rangle 
  &=&  \frac{ \langle \psi | {\cal O}^\dagger_i|\psi \rangle }
      {|\langle \psi | \psi \rangle|^2 }
      = \frac{ \sum_m\langle \psi | {\cal O}^\dagger_i|m\rangle \langle m|\psi \rangle }
      {\sum_m |\langle m | \psi \rangle|^2 }      
 \nonumber \\
  &=&  \frac{ \sum_m|\langle \psi |m \rangle|^2
    O^*_i(m, \alpha) }
  {\sum_m |\langle m | \psi \rangle|^2 }
  \nonumber \\
  &=& \sum_m p(m) O^*_i(m, \alpha) 
\end{eqnarray}
where $p(m)$ is defined as 
$p(m)=|\langle m|\psi\rangle|^2 / \sum_{m'}|\langle m'|\psi\rangle|^2 $. 
The weighted summation $\sum_m p(m)$  is 
realized by the Markov Chain Monte Carlo (MCMC) process
in which $|m\rangle$ is generated obeying the probability $p(m)$.
The energy is also evaluated in the same manner as
\begin{eqnarray}
  E_L(m) &=& \frac{\langle m | H |  \psi \rangle }{\langle m | \psi \rangle }
             \nonumber \\
  \langle H \rangle &=& \sum_m p(m) E_l(m)
\end{eqnarray}
Other relevant values are evaluated as 
\begin{eqnarray}
  \langle {\cal O}_i \rangle 
  &=&   \sum_m p(m) 
  O_i(m, \alpha)  =  \langle {\cal O}^\dagger_i \rangle^*
\end{eqnarray}

\begin{eqnarray}
  \langle {\cal O}^\dagger_i {\cal O}_j \rangle 
  &=& \sum_m p(m) 
      O^*_i(m, \alpha)  O_j(m, \alpha),
\end{eqnarray}

\begin{eqnarray}
  \langle {\cal O}^\dagger_i H \rangle 
  &=&  \frac{ \sum_{m}\langle \psi | {\cal O}^\dagger_i|m\rangle
    \langle m|H | \psi \rangle }
  {\sum_m |\langle m | \psi \rangle|^2 }
  \nonumber \\
   &=& \sum_m p(m) 
   O^*_i(m, \alpha) E_L(m),  
\end{eqnarray}
\begin{eqnarray}
  \langle  H  {\cal O}_i\rangle 
  &=& \sum_m p(m) 
  E_L^*(m) O_i(m, \alpha) .
\end{eqnarray}

The derivative concerning the operator $G$ 
is evaluated as
\begin{eqnarray}
  O_{\alpha_{ij}}(m,\alpha) &=& \frac{1}{\langle m| \psi \rangle }
  \frac{\partial}{\partial \alpha_{ij}}
  \langle m| e^{\sum_{i\leq j} \alpha_{ij} n_{i} n_{j}}| \psi\rangle
  \nonumber \\
  &=& \sum_{i\leq j}  n^{(m)}_{i}   n^{(m)}_{j} 
\end{eqnarray}
with $ n^{(m)}_{i} = \langle m|n_{i}|m \rangle $.

The derivative concerning correlated pairs is 
\begin{eqnarray}
  && O_{f_{ij}}(m,\alpha) 
  \nonumber \\
  &=& \frac{1}{\langle m | \psi\rangle}
  \frac{\partial}{\partial (f_m)_{ij}}\langle m | \psi\rangle
  \nonumber \\
  &=& \frac{1}{\gamma_m 2^{N/2}(N/2)! {\rm Pf}(f_m)}
  (- (f_m)^{-1}_{ij} {\rm Pf}(f_m)\gamma_m 2^{N/2}(N/2)! )
  \nonumber \\
  &=& - (f_m)^{-1}_{ij} = - f_{m_i m_j}
  \nonumber \\
  &=&  \frac12 \left( (f_m)^{-1}_{ji} - (f_m)^{-1}_{ij} \right)  
\end{eqnarray}

The derivative concerning the 
correlated-pair parameters of 
the $J$-projected energy is 
\begin{eqnarray}
  && O_{f_{kk'}}(m,\alpha) 
  \nonumber \\
  &=& \frac{1}{\langle m | P^J_M| \phi\rangle}
  \frac{\partial}{\partial X_{ab}}\langle m | P^J_M |\phi\rangle
  \\
  &=& \frac{1}{\sum_{nK} g_K w_{nK} \langle m |R_n | \phi\rangle}
  \sum_{nK} g_K w_{nK} \frac{\partial}{\partial X_{ab}}\langle m |R_n | \phi\rangle
  \nonumber \\
  &=& \frac{1}{\sum_{nK} g_K w_{nK}  \langle m |R_n | \phi\rangle}
  \sum_{nK} g_K w_{nK}  \langle m |R_n | \phi\rangle
  \nonumber \\ 
  &&
  \times \left(- \sum_{i,j=1}^{N} R_{am_i}^T ((RXR^T)_m)^{-1})_{m_im_j} R_{m_jb} \right)
  \nonumber
\end{eqnarray}

The derivative concerning the $g_K$ is 
\begin{eqnarray}
  && O_{g_K}(m, \alpha) 
  \nonumber \\
  &=& \frac{1}{\langle m | \psi\rangle}
  \frac{\partial}{\partial g_K}\langle m | \psi\rangle  
  \\
  &=& \frac{1}{\sum_{nK'} g_{K'} w_{nK'} \langle m | R_n | \phi \rangle} 
  \sum_{n}  w_{nK} \langle m | R_n | \phi \rangle
   \nonumber 
\end{eqnarray}
By combining these equations and the MCMC procedure, 
we can evaluate the energy gradient 
of the $J$-projected energy.


The norm of the small displacement of the $|\overline{\psi}\rangle $ 
caused by the small change of the variational parameters $\gamma_i$ is
\begin{eqnarray}
  \Delta^2_{\rm norm} & = & \left|\left| |\overline{\psi}_{\bf \alpha + \gamma}\rangle
      - |\overline{\psi}_{\bf \alpha }\rangle  \right|\right|^2
  \nonumber \\
  &=& \sum_{ij} \gamma_i^* \gamma_j 
      \frac{\partial}{\partial \alpha^*_i}  
      \frac{\partial}{\partial \alpha_j}  
      \langle \overline{\psi}|\overline{\psi}\rangle 
  \nonumber \\
  &=& \sum_{ij} \gamma_i^* S_{ij} \gamma_j 
\end{eqnarray}
with the overlap matrix $S_{ij}$, 
\begin{eqnarray}
  S_{ij} 
  &=& \langle {\cal O}^\dagger_i {\cal O}_j \rangle 
      - \langle {\cal O}^\dagger_i \rangle \langle {\cal O}_j \rangle 
\end{eqnarray}
which is Hermitian and positive semidefinite \cite{nnqs-prx}.

In the steepest gradient method, 
the small displacement is taken as the derivative of energy 
such as 
\begin{equation}
  \gamma_i = - \Delta t \frac{\partial \langle H \rangle}{\partial \alpha^*_i} 
  = - \Delta t g_i . 
\end{equation}
On the other hand, in the SR method, 
the small displacement is taken as the
the product of the inverse of $S_{ij}$  and  derivative of energy such as
\begin{eqnarray}
  g'_i &=&  - \Delta t \sum_j  S_{ij}^{-1} g_j 
  \label{eq:gradSR}
\end{eqnarray}
By using the inverse of  $S_{ij}$,
the direction with the small norm of the $S_{ij}$, 
or the direction causing small displacement, 
is taken as large step width and vice versa.
In this work, we typically take $\Delta t = 0.2$. 

In order to stabilize the SR method further, 
we apply two modifications to the overlap matrix, $S_{ij}$
following Ref.~\cite{Tahara}.
One is the scaling of its diagonal matrix elements. 
We replace the overlap matrix by the scaled one, 
\begin{equation}
  S'_{ij}  = (1 + \epsilon \delta_{ij}) S_{ij} , 
\end{equation}
where $\epsilon$ is a small constant.
This modification makes the overlap matrix positive definite 
and stable
even if $S_{ij}$ is calculated stochastically including a certain error 
\cite{sorella_scalediag}. 
In this work, we typically take $\epsilon=0.01/\sqrt{i}$  
where $i$ is the number of iterations.

The other method to stabilize the SR method 
is the truncation of the redundant directions by introducing 
the cut off of the small eigenvalues of the overlap matrix.
As it is Hermitian, we can diagonalize the overlap matrix such as
\begin{equation}
  S_{ij} = \sum_k U_{ik} \lambda_k U^\dagger_{kj}. 
\end{equation}
The redundancy of the variational-parameter space causes zero or small
eigenvalues of the overlap matrix.
Besides, small eigenvalues with statistical errors cause
instability in evaluating the inverse matrix in Eq.(\ref{eq:gradSR}).
In order to avoid the problem, we replace $1/\lambda_i$ by 0 for 
$\lambda_i<\epsilon_{\rm cut}$. 
In this work, we typically take 
$\epsilon_{\rm cut} = 2/\sqrt{i}\times 10^{-4}$ where $i$ is the number 
of iterations. Thus, 
\begin{equation}
  \gamma_k = -\Delta t  \sum_lS^{-1}_{kl} g_l  =
  -\Delta t \sum_{il} \frac{1}{\lambda_i} U_{ki} U^\dagger_{il}  g_l 
\end{equation}
is replaced by 
\begin{equation}
  \gamma_k = 
  -\Delta t \sum_{il}
    \Theta(\lambda_i-\epsilon_{\rm cut})
    \frac{1}{\lambda_i} U_{ki} U^\dagger_{il}  g_l 
\end{equation}
where $\Theta(x)$ is the Heaviside function.

As a summary, we iteratively shift the variational parameters 
by adding the direction provided by Eq.~(\ref{eq:gradSR}) 
in the SR method.
It is expected to decrease the energy expectation value 
and, at the same time, to suppress the norm of 
the displacement of the wave functions 
by removing the effect of the redundancy of the variational parameters.
This procedure is iterated until the energy converges.


\end{document}